# High Resolution Image Reconstruction Method for a Double-plane PET System with Changeable Spacing*

GU Xiao-Yue (顾笑悦)[1;2] ZHOU Wei(周魏)[1;2] LI Lin (李琳)[1;2] YIN Peng-Fei (尹鹏飞)[1;2] SHANG Lei-Min(尚雷敏)[1;2] YUN Ming-Kai (贠明凯)[1;2] LU Zhen-Rui(卢贞瑞)[1;2] HUANG Xian-Chao (黄先超)[1;2] WEI Long(魏龙)[1;2]

1 (Key Laboratory of Nuclear Radiation and Nuclear Energy Technology, Institute of High Energy Physics, Chinese Academy of Sciences，Beijing 100049, China)

2 (Beijing Engineering Research Center of Radiographic Techniques and Equipment, Beijing 100049, China)

**Abstract:** Positron Emission Mammography (PEM) imaging systems with the ability in detection of millimeter-sized tumors were developed in recent years. And some of them have been well used in clinical applications. In consideration of biopsy application, a double-plane detector configuration is practical for the convenience of breast immobilization. However, the serious blurring effect in the double-plane system with changeable spacing for different breast size should be studied. **Methods:** We study a high resolution reconstruction method applicable for a double-plane PET system with a changeable detector spacing. Geometric and blurring components should be calculated at real time for different detector distance. Accurate geometric sensitivity is obtained with a tube area model. Resolution recovery is achieved by estimating blurring effects derived from simulated single gamma response information. **Results:** The results show that the new geometric modeling gives a more finite and smooth sensitivity weight in double-plane system. The blurring component yields contrast recovery levels that could not be reached without blurring modeling, as well as better visual recovery of the smallest spheres and better delineation of the structures in the reconstructed images. Statistical noise has lower variance at the voxel level with blurring modeling than without at matched resolution. **Conclusion:** In the distance-changeable double-plane PET, the finite resolution modeling during reconstruction achieves resolution recovery, without noise amplification.

## 1 Introduction

Breast-dedicated PET imaging systems have been developed in recent years. And the systems' abilities in detection of millimeter-sized breast tumors were studied [1-10]. Some of them have been demonstrated clinically feasible and valuable in detection of breast tumors [4, 7, 11, 12]. What's more, better spatial resolution than whole-body PET was reported for most of the systems. The two main recent fashions in breast-dedicated PET are ring and plane detector configurations.

In consideration of biopsy application, a double-plane detector configuration is more practical than a ring detector for the convenience of breast immobilization. And a double-plane PET system with a small spacing could achieve a higher sensitivity. With a small distance between the double-plane PET, the parallax errors caused by the oblique incidence gamma rays could result in the deterioration of the image qualities [13]. The penetration of the 511 keV photons into the crystals are severer when a photon incident on the detector with a larger oblique angle into the crystal faces. The effect will deteriorate resolution as well as offset the advantages gained in the sensitivity.

A number of hardware approaches have been proposed to compensate this effect. These methods are capable of providing depth-of-interaction (DOI) information [14-16]. But the complexity of detector system design, the accuracy of measurements, and the cost require further investigation. Nevertheless, there is another approach to compensate the parallax errors by establishing an accurate system response matrix in reconstruction [13, 17, 18]. The improvement with the quality of reconstructed images depends on an accurate model of the relationship between image and projection space [19, 20]. In general, geometrical component and detection physics effects information, or blurring factor, of the system matrix should be well investigated in the resolution model (RM).

The geometric elements of the system matrix could be calculated by using simple line integral model [21].

___
Received date 16 July 2015
*Supported by Knowledge Innovation Project of The Chinese Academy of Sciences (KJCX2-EW-N06)
1) E-mail: guxiaoyue@ihep.ac.cn

More complex model takes LOR as tube [9, 20, 22] or based on solid angle [23]. To achieve a feasible reconstruction times, pre-calculation and storage of the system matrix are always required in these cases, commonly with the geometrical symmetries used. In advance, improvement in spatial resolution can be achieved by modeling detector blurring effects including crystal penetration, inter-crystal scattering and crystal misidentification [19, 24-26].

PEM Flex Solo scanner was the first commercial machine produced by Naviscan Inc with a double-plane PET configuration. The 6cm × 16.4cm detectors are positioned in opposing fashion and should move to cover the whole 24cm × 16.4cm field of view (FOV) [6]. The correction method of parallax errors was not reported in the study of Naviscan. Note that the changeable distance for different breast size in a plane geometric make this effect more difficult to deal with. Reconstruction method for double-plane detector were studied by Chien-Min Kao. Resolution recovery was achieved and the system response matrix was drastically reduced in their work [13]. However, the simulations work was applicable for a system with static plane spacing. When the distance changed, the SRM simulation should be repeated for the new detector spacing. Therefore the method is not suitable for the breast imaging application.

The demand for a new reconstruction method with resolution modeling (RM) is motivated by the need to develop a high resolution and sensitivity double-plane system with a changeable spacing. The system designed with a final target to well perform biopsy of the breast. And the distance between the planes would be changed as size of the breast. We focus our method on geometrical component and the blurring effect in the double-plane system. The RM method is a combination of Monte Carlo simulation and calculation solution [27].

## 2 Materials and Methods
### 2.1 Reconstruction for double-plane PET

In this study, we present a new RM method for the resolution recovery in the double-plane system. The system is supposed as a PET system composed of two opposing detector heads as in figure 1, with a changeable distance d between PET plans for different breast size. The orientations of the cross-plane (blue plane in figure 1) and in-plane (red plane in figure 1) directions are also showed. In general, statistical reconstruction methods can include the factors in the system matrix that represents the probability of detecting an emission from each image voxel at each detector-pair. And resolution modeling is carried out in projection space with 3D EM algorithm in consideration.

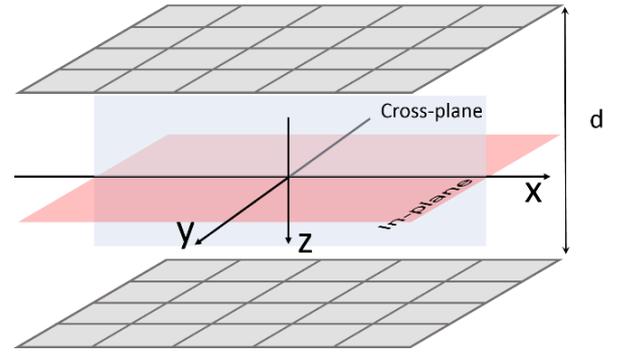

Fig. 1. The PET system composed of two opposing detector heads. The distance d between the planes is changeable for different breast size. The orientations of the in-plane and cross-plane directions are showed in the figure.

Of first importance to any implementation of an iterative reconstruction algorithm is the system model. We denote the system matrix as $\mathbf{P} \in R^{J \times I}$ whose elements $p_{ij}$ model the probability that an event generated in voxel **j (j=1···J)** is detected along a LOR **i (i=1···I)**. The system matrix is factorized as follows:

$$P = P_{det.sens} P_{det.blur} P_{attn} P_{geom} P_{positron}. \quad (1)$$

Here $\mathbf{P}_{attn} \in R^{I \times I}$ is a diagonal matrix containing the attenuation factors. We applied a calculated attenuation correction method based on breast image segmentation. Attenuation factors are obtained from re-projecting of the estimated attenuation map [28]. The diagonal detector normalization matrix $\mathbf{P}_{det.sens} \in R^{I \times I}$ is taken as uniform for the simulated data generated from identical crystals. However, we therefore simplify the model with the 18F application and lump the positron range effect $\mathbf{P}_{positron}$.

In traditional PET systems, the elements $p_{ij}$ numbers of system matrix **P** is always a constant. However, the changeable distance property makes the size of the image space is variable in the double-plane system. With a fixed voxel size, the voxel number **J** would change with the detect spacing. Therefore, the factors $\mathbf{P}_{det.blur}$ and $\mathbf{P}_{geom}$ should be calculated at real time in every scan. With this new attribution, we focus on the two factors in our model.

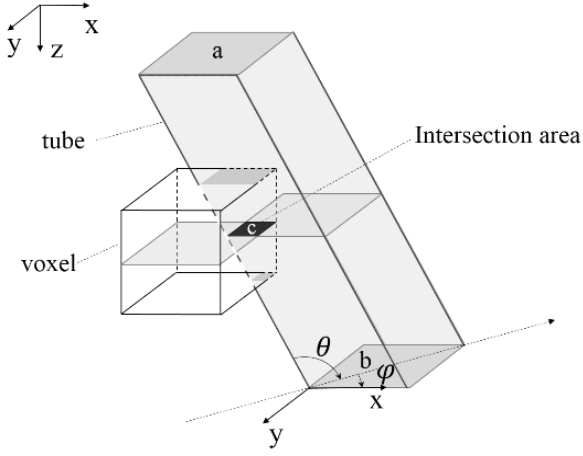

Fig. 2. The relationship of a voxel and tube.

### 2.1.1 The Geometric projection matrix

$P_{geom} \in R^{J \times I}$ is a matrix that contains the geometrical mapping between the source and sino data. Each element **(i,j)** of $P_{geom} \in R^{J \times I}$ represents the probability that a photon pair produced in voxel **j** reaches the front faces of the detector pair i. The tube model is taken to compute the intersection joining the detector pair with each voxel. And a finite weight factor is applied based on the geometric property of the plane PET.

The image coordinate system is defined as Cartesion **(x, y, z)**. When the double planes are extremely in parallel, the middle plane of the voxel section is paralleled with the in-plane direction. Figure 2 illustrated the relationship of a voxel and tube in the double-plane system. The intersection of the tube with the center plane of voxel is marked as dark gray. Note that the upper detector face *a* is in parallel with the intersection area *c*. And the intersection area *c* (see figure 3) with the LOR tube keeps a rectangle shape for every element **(i,j)** of the $P_{geom}$. The weight value of element **(i,j)** can be parameterized by the area **a(i,j)** of rectangle *c*, over which the area value is easy calculated in a parallel plane PET by defining the four side boundaries of the rectangle. The formula (2)-(6) illustrate the boundary calculation in image slice $z_0$.

$$a(i,j) = \left(L_{right}(i,j) - L_{left}(i,j)\right) \times \left(L_{down}(i,j) - L_{top}(i,j)\right). \quad (2)$$

$$L_{left}(i,j) = \max\left(x_{voxel,left}(z_0), x_{tube,left}(z_0)\right). \quad (3)$$

$$L_{right}(i,j) = \min\left(x_{voxel,right}(z_0), x_{tube,right}(z_0)\right). \quad (4)$$

$$L_{top}(i,j) = \max\left(y_{voxel,top}(z_0), y_{tube,top}(z_0)\right). \quad (5)$$

$$L_{down}(i,j) = \min\left(y_{voxel,down}(z_0), y_{tube,down}(z_0)\right). \quad (6)$$

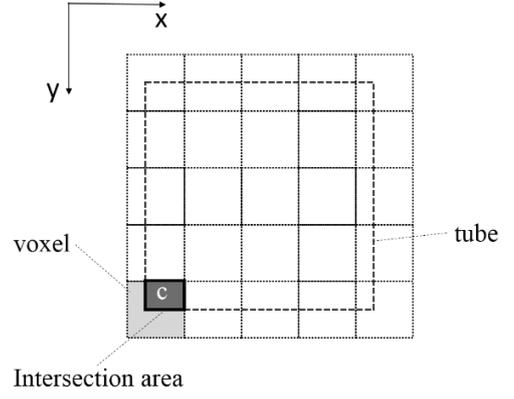

Fig. 3. The model of the finite weight. In the parallel double-plane system, the intersection area illustrated in the figure is proportional to the intersection area of the tube and the voxel.

The factor **a(i,j)** is the area value of the intersection area *c*. $L_{left}(i,j)$, $L_{right}(i,j)$, $L_{top}(i,j)$, $L_{down}(i,j)$ respectively represents the four side boundaries of area *c*. The four boundaries are all obtained by comparing the corresponding boundaries of voxel and tube in in-plane as in figure 3. For example, the left boundary $L_{left}(i,j)$ in formula (2) is the larger **x** value between the left boundaries of the voxel and the tube.

Suppose the detector element is 2mm × 2mm and the image voxel is 0.5mm × 0.5mm in in-plane, the length of the tube side is four times of the voxel and the intersection area *c* is illustrated as in figure 3. We take the intersection area as the finite weight of the value of the $P_{geom}$ element. In the parallel double-plane system, the intersection area is proportional to the intersection volume of the tube and the voxel.

### 2.1.2 The sinogram blurring matrix

$P_{det.blur}$ means the sinogram blurring matrix used to model photon inter-crystal penetration and inter-crystal scatter affect. We treat each crystal as identical and therefore ignore effects associated with the location within each block [25]. In principle, the non-collinearity of the photon pair should be taken into consideration. However, we therefore simplify the model and lump the angle separation effect in the small detector. We also did not include subject scattering or positron range in the simulation work.

The detection planes are parallel to the in-plane, located at positions $z = \pm d/2$ (see figure 1). We note that double-plane PET projection data are described as $(\vec{C_u}, \vec{C_l})$. $\vec{C_u}$ and $\vec{C_l}$ denote the indented crystal element in the upper and lower detection planes respectively. We use $(\overrightarrow{C_u}, \overrightarrow{C_l}, \overrightarrow{C_{u'}}, \overrightarrow{C_{l'}})$ maps a given $\mathbf{LOR}\,(\overrightarrow{C_u}, \overrightarrow{C_l})$ to its blurred counterparts $\mathbf{LOR'}(\overrightarrow{C_{u'}}, \overrightarrow{C_{l'}})$. We assume that the detector tube is rotated in two directions by oblique angle of $\varphi$ and azimuthal angle of $\theta$. Note that $(\overrightarrow{C_u}, \overrightarrow{C_l}, \overrightarrow{C_{u'}}, \overrightarrow{C_{l'}})$ could also be expressed as $(\overrightarrow{C_u}, \overrightarrow{C_l}, \varphi, \theta)$ with the LOR rotation angle $\varphi$ and $\theta$ defined.

The distance of source-voxel to detector is demonstrated has an effect in the presence of axial mashing (spanning) in a ring configure detector [19, 25, 29]. Nevertheless, how the distance of source-voxel affect distribution of penetration in the double-plane PET is studied. Back to back gamma ray source at different distance positions to the detector face was simulated with specified incidence angle ($\varphi = 0$ deg, $\theta = 45$ deg) as illustrated in figure 4(a). The detector spacing is 2cm and the source voxel position ranges from center to the surface of the lower detector plane with a step of 1mm, along the LOR direction. The coincidence response of crystal $\mathbf{u}$ with each crystal element of the below detector was studied, says crystal $\mathbf{l}$, $\mathbf{l'}$, $\mathbf{l''}$ etc. And the profile of the response $\mathbf{LOR(u, l)}$ with its main blurred counterparts in the lower plane $\mathbf{LOR'(u, l')}$ and $\mathbf{LOR''(u, l'')}$ was plotted in figure 4(b). The profile illustrates that different source positions results turn as consistent response distribution. And the distribution of the incidence light is independent of the source position in a double-plane PET system.

Based on the above result, we assume that the response of gamma ray blurring effects in the plane PET is independent of the source position and mainly affected by the incidence angle direction into the crystal. In our current implementation we approximate the blurring effects as the probability product of two separated single gamma ray penetrated effects. In summary, the general blurring function expressed as coincidence response function(CRF) $(\overrightarrow{C_u}, \overrightarrow{C_l}, \overrightarrow{C_{u'}}, \overrightarrow{C_{l'}})$, which could be described as follows.

$$\begin{aligned}\mathbf{CRF}(\overrightarrow{C_u}, \overrightarrow{C_l}, \overrightarrow{C_{u'}}, \overrightarrow{C_{l'}}) \\ = \mathbf{CRF}(\overrightarrow{C_u}, \overrightarrow{C_l}, \varphi, \theta) \\ = \mathbf{SGR_u}(\overrightarrow{C_u}, \overrightarrow{C_{u'}}, \varphi, \theta) \times \mathbf{SGR_l}(\overrightarrow{C_l}, \overrightarrow{C_{l'}}, \varphi, \theta).\end{aligned} \quad (7)$$

The single gamma response(SGR) $\mathbf{SGR}(\overrightarrow{C_u}, \overrightarrow{C_{u'}}, \varphi, \theta)$ function represents the crystal $\mathbf{u'}$ response probability when the gamma ray incidence to crystal $\mathbf{u}$, with an oblique angle of $\varphi$ and azimuthal angle of $\theta$.

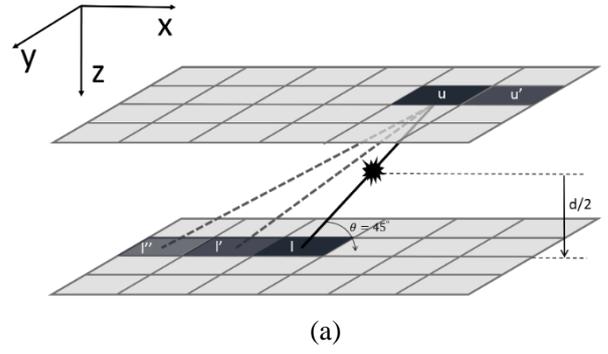

(a)

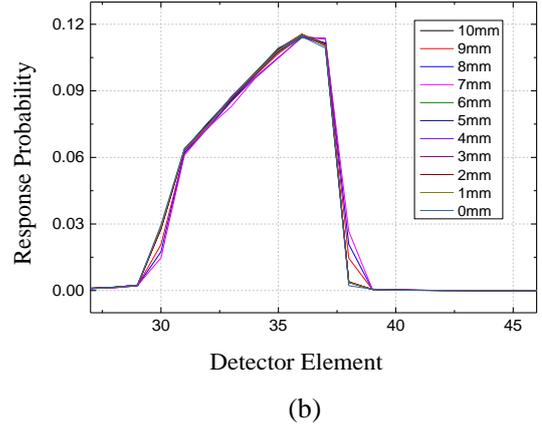

(b)

Fig. 4. Back to back gamma ray source with different distance to the detector was simulated in certain incidence angle ($\varphi=0$ deg, $\theta=45$ deg)

The SGR was modelled along two dimensions crystal arrays with different incidence angle. In advance, the SGR simulated work could be effectively reduced in consider of the plane system symmetry of angle $\varphi$. The symmetry is as follows:

$$\mathbf{SGR}(\overrightarrow{C_u}, \overrightarrow{C_{u_0'}}, \varphi, \theta) = \mathbf{SGR}(\overrightarrow{C_u}, \overrightarrow{C_{u_1'}}, 90°+\varphi, \theta),$$
$$u_0'(x,y) = u_1'(y, -x) \quad (8)$$

$$\mathbf{SGR}(\overrightarrow{C_u}, \overrightarrow{C_{u_0'}}, \varphi, \theta) = \mathbf{SGR}(\overrightarrow{C_u}, \overrightarrow{C_{u_2'}}, 180°+\varphi, \theta),$$
$$u_0'(x,y) = u_2'(-x, -y) \quad (9)$$

$$\mathbf{SGR}(\overrightarrow{C_u}, \overrightarrow{C_{u_0'}}, \varphi, \theta) = \mathbf{SGR}(\overrightarrow{C_u}, \overrightarrow{C_{u_3'}}, 270°+\varphi, \theta),$$
$$u_0'(x,y) = u_3'(-y, x) \quad (10)$$

With the reduction, the simulation work of SRF was effectively reduced into 1/4 with angle $\varphi$ span from 0 to 90 degree. With the parallel attribute of the double-plane, the upper detector $\mathbf{SGR_u}(\overrightarrow{C_u}, \overrightarrow{C_{u'}}, \varphi, \theta)$ and lower detector $\mathbf{SGR_l}(\overrightarrow{C_l}, \overrightarrow{C_{l'}}, \varphi, \theta)$ has a symmetric about the origin.

The simulation work is developed on the work by Fan xin [30] and extended into 3D implementation. The single photon incidence response was obtained with Monte Carlo simulation. Geant4 method for emission tomography (GATE) software was applied. A two-dimensional crystal array with 33 $\times$ 33 crystal elements was created as in figure

5 (only 5 × 5 elements illustrated in the figure). The simulated crystal size is 1.9 mm × 1.9 mm × 10mm with a 0.1mm gap filled with polyvinyl chloride (PVC), which is the same set as the double-planes system. The simulated work was performed by rotating the single detector along the **x** and **y** axis with two directions **θ** and **ψ**. The single gamma ray was incidence into the center crystal **u**. The two directions both span from 0 to 90 degree cover with all the incidence angle span for different plane distance. There were total 18 × 18 (324) directions to simulate with a five-degree step.

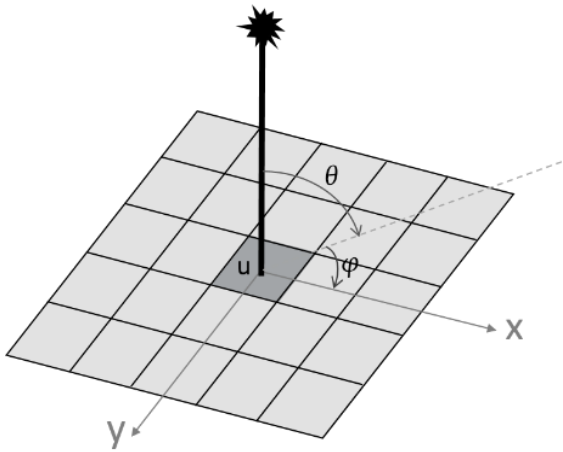

Fig. 5.    Single gamma response simulation.

The simulated single gamma ray response results of three incidence angles were showed in figure 6. Plots from (a) to (c) illustrated the signal gamma response of the 30, 45, 60 degree respectively. The response result turned as a distribution of the event counts collected within the crystals. In most condition, the reaction happened in the gamma ray trajectory while most crystal out of the trajectory turned out with a little events count. The finally blurring factor would be calculated based on single gamma ray response using formula (7). The simulated SRG is discrete with five-degree span. The certain SRG with an identified incidence angle calculated from the simulated SRG by linear interpolation.

The calculated CRF derived from SGR was tested. Three incidence angles were chosen to compare the real CRF with the calculated CRF results. The calculated value was derived from the product of $SGR_u(\vec{C_u},\vec{C_{u'}},\varphi,\theta) \times SGR_l(\vec{C_l},\vec{C_{l'}},\varphi,\theta)$. The real CRF was simulated and obtained with MC method. The incidence angles were chosen as $\varphi = 0$ deg, $\theta = 30, 45, 60$ deg for the convenience for simulate. Figure 7 shows that the compare results of the real and calculated CRF. The crystal chosen and data selection is similar as in the source position study. The profile shows that calculated CRF (upper profile in figure 7) is a good approximation with the real CRF (lower profile in figure 7).

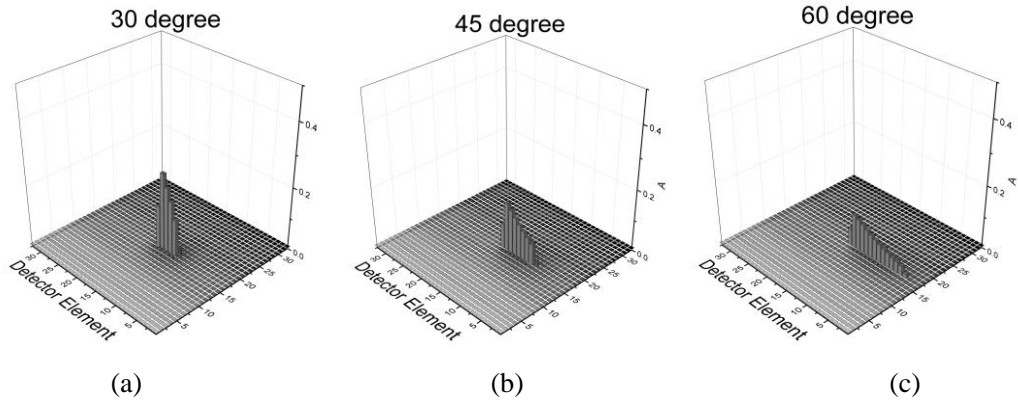

(a) (b) (c)

Fig. 6.    Simulated single gamma ray response results with three incidence θ angles. (a) θ = 30 deg. (b) θ = 45 deg. (c) θ = 60 deg.

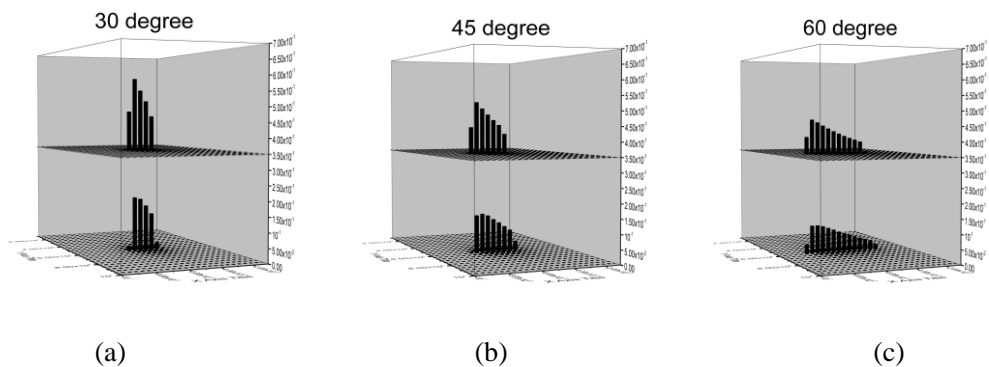

(a) (b) (c)

Fig. 7.    Coincidence response function of LOR. Upper slice is calculated CRF and lower slice is real CRF. (a) θ=30 deg. (b) θ=45 deg. (c) θ=60 deg.

## 2.2 PET Implementation of the Algorithm

To evaluate the algorithm, a PET system composed of two opposing detector heads was simulated. The distance between two detector heads ranged from 1cm to 6cm, assuming as the distance range for breast imaging application. Both detectors contains 75 × 100 LYSO crystal elements. The crystal size is 1.9 mm × 1.9 mm × 10mm with a 0.1mm gap filled with PVC material. Note that the SGR simulated data generated from the same crystal configuration as the opposing detector system but with a proper crystal element number. Data acquisition and reconstructed performed in 3D. The algorithm operation is described by matrix P in Equations 1-7. The reconstructed images matrix size in in-plane were 400 ×300 with a pixel size of 0.5 mm. In the axial direction the image pixel size is 1mm. The image slice number is defined by the detector spacing d. The geometric and blurring factors were both assessed. The reconstructions was accelerated with OpenMP parallel programming support. The reconstruction using the blurring factor modeled system matrix will be referred as the resolution modelling (RM) reconstruction.

To assess the impact of the geometric components on the image quality. A cube (16 pixels side length) and a sphere (16 pixels diameter) source with 18F was placed in air. Two shapes were both with 5000Bq/ml activity concentration. Reconstructions without blurring factor and tested with three different geometric weights, respectively ray-driven model[21], solid weight model[20] and the new tube area model. The sino data were reconstructed with EM (30 iterations, no subsets). All span data in sinogram were used.

A numerical micro-Derenzo phantom was used to generate noise-free data and to test the blurring factor. The diameters of the hot spots in the phantom are 2.4 mm, 2.0 mm, 1.7 mm, 1.35 mm, 1.0 mm, and 0.75 mm respectively and the center-to-center distance between the spots is twice the hot spot diameter. All the hot spots with an activity concentration of 5000Bq/ml. The height of phantom was 10mm with the axis vertical to the in-plane direction. The phantom was located at center of the FOV. The distance between two detector heads ranged from 1cm to 6cm with the reconstructed image slices ranged from 10 to 60. Data were reconstructed by the 3D MLEM, respectively with (RM) or without (no-RM) the blurring information in the SRM. Only the true coincidence events was used. We choose the image reconstructed after 15 iterations and plotted the images in the center in-plane slice.

The noise properties were evaluated with a cylinder contrast phantom (cylinder length 10mm, radius 27mm). This phantom consists of five hot spherical inserts of decreasing size, containing a uniform background activity concentration of 5000Bq/ml. The embedded five hot spheres with the radius range from 1mm to 5mm. All the hot spheres have a 4:1 ratio to background activity. The phantom is placed at center of FOV and the detector distance is 2cm. Contrast recovery and noise characteristics are investigated between the RM and no-RM algorithms. The contrast ratio is tested for each sphere i with by the mean signal for each sphere $S_i$ against the background $B_i$. Background volumes of interest (VOIs) of the same volume as the spheres are chosen between different in-planes at each sphere's (x,y) locations.

For each sphere, the contrast ratio for each sphere are as follows:

$$\text{contrast ratio} = \frac{\langle S_i \rangle}{\langle B_i \rangle} \qquad (11)$$

For each sphere, the background **SN** for each sphere were then found using:

$$SN = \frac{std(B_i)}{\langle B_i \rangle} \qquad (12)$$

Where <> represents the mean and std() is the SDs across all pixels.

## 3 Results
### 3.1 Geometric component

The comparison of the results (Fig. 8) respectively with ray-driven model[21], solid weight model[20] and the new tube area model shows the dramatic improvement obtained with the latter. The tube area model obtains a relatively smooth result, while artifact errors are showed in the other two models.

### 3.2 Blurring component

Figure 9 shows the reconstructed image of the micro-Derenzo phantom. The hot rods with a diameter of 1.35 mm can be identified in both RM and no-RM data. And the structure of the region containing 1.0 mm diameter rods can also be observed in the data with RM data while the same size hot rods are blurred in the no-RM data. RM data shows a better visual recovery of the smallest spheres and better delineation of the structures in the reconstructed images are observed. No-RM data show an increased blurring effects as the detector distance become smaller. Results show the profiles of the third line hot spots in the micro-Derenzo phantom. Overall, contrast of the hot rods against background remained identical with RM compared with no-RM results. Graph of the selected three group hot rods demonstrate that slimmer profiles are observed when RM is used, leading to lower spatial variance. All the images are showing with the same viewing window width but slightly different viewing means.

### 3.3 Quality study

Figure 10 shows the impact of the RM on reconstruction of the noise properties. The contrast ratio plot shows background contrast was improved with RM data for each sphere size. However the background noise ratio are decreased with each VOIs for all the the sphere size. It is confirmed that RM significantly reduced the voxel variance to a level comparable with the level that was obtained after reconstruction without RM. Figure demonstrates that higher positive correlations with adjacent voxels are observed when RM is used, leading to lower spatial variance.

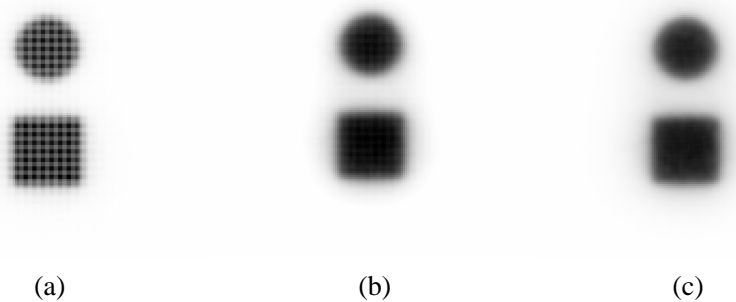

(a) (b) (c)

Fig. 8. Reconstructions of sources data for different geometric models. (a) Ray-driven model. (b) Solid angle model. (c) Tube area model.

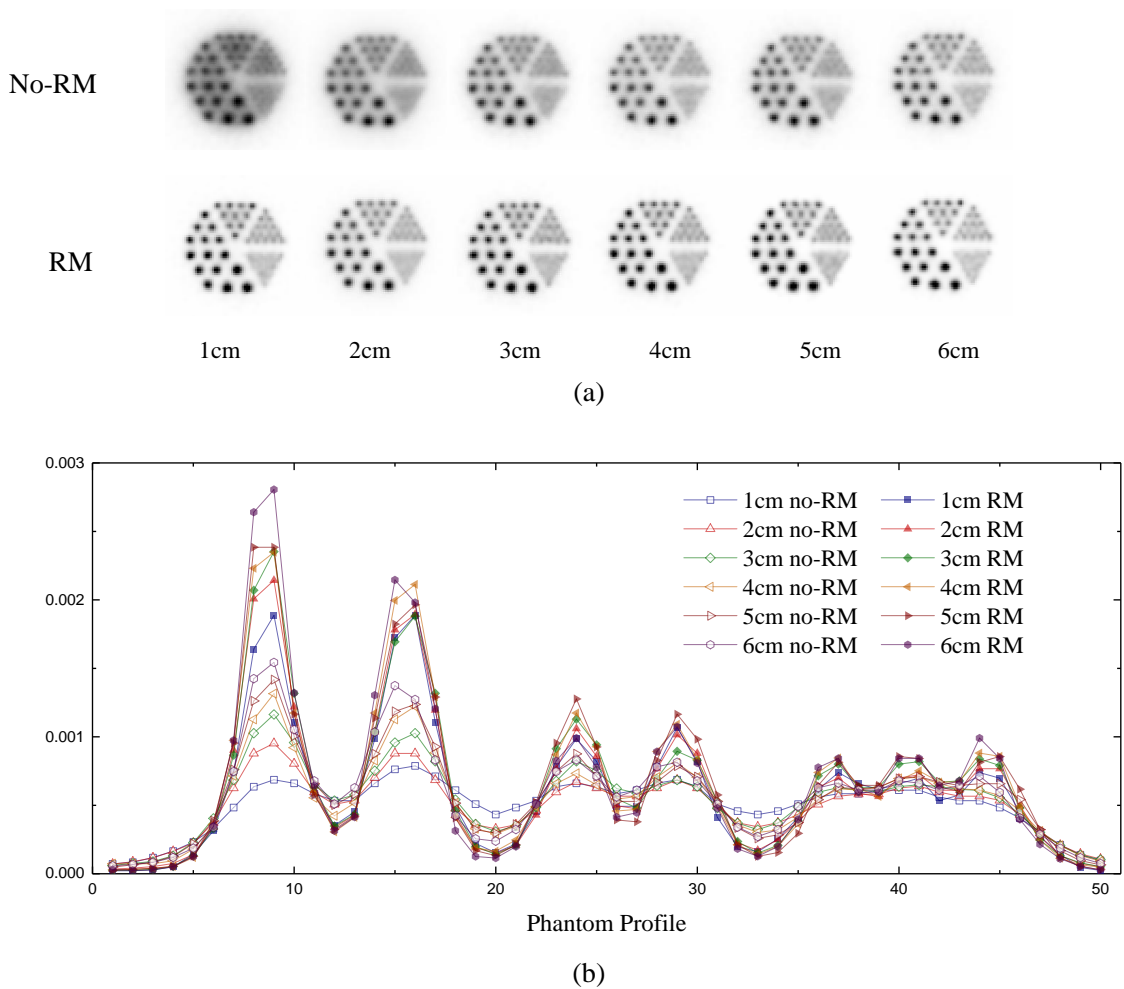

Fig. 9. Micro-Derenzo phantom results.

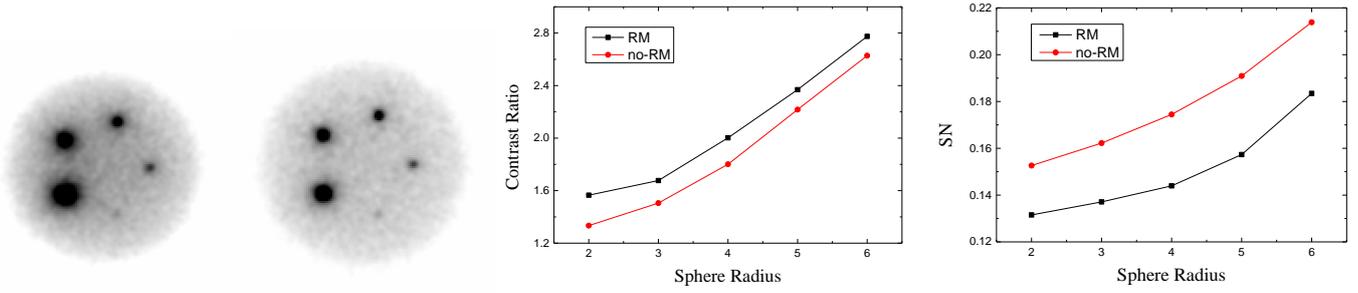

Fig. 10. Quality phantom results.

## 4 Discussion

The results demonstrates that RM in the reconstruction process improves spatial resolution (giving better delineation of the structures), improves contrast recovery, and also improves the noise properties of the images. The method has following features: (1) it is applicable for different plane spacing; (2) it is efficient in simulation; (3) it is effective for resolution recovery; and (4) it is robust in noise suppression.

When employing double-plane geometry, the artifacts result from missing data make the existing analytic reconstruction methods will not work well. In contrast, iterative reconstruction algorithms that are based on a statistical model are able to maximize resolution recovery by accurate modelling of the system response. Furthermore, iterative reconstruction methods can optimize performance in low-count situations. For the listed concerns, the iterative method is applied in this study. The accurate system modelling is focused on geometric and blurring components.

It is interesting to note that the artifacts in the figure 8(a) and (b) are grids shape in both in **x** and **y** axial directions. The errors were generated probably due to the discontinuous "square pixel" modeling of the reconstructed image. The discrete contribute could be compensated in a full ring detector with the oversampling data in transaxial direction. However, the effect could not be ignored in the plane system with the paralleled sampling in in-plane while the data are missing in the cross-plane. The algorithm, based on intersection area between voxel and tube, provides superior accuracy and smooth geometric sensitivity weights without loss of resolution. Furthermore, the tube area model takes the advantages of the parallel attributes of the image space and it is applicable for the double PET system.

The blurring component is based on a single gamma response calculated model. The simulated single gamma response matrix size is a trade-off between the precision and computation boundary. The 33 crystal number in SRG simulation is a compromise for the computation boundary. When the incidence angle is extremely oblique, there is a precision lost for the tail-cut because of the limited simulated crystal array. However, the precision lost is less obvious in the reconstructed result as showed in the micro-Derenzo phantom and quality phantom. In advance, a simulated matrix reduction to obtain a fast computation rate is in further study.

The system operated as the traditional mammography could results in substantial parallax errors. The effects could be observed from figure 7. When the distance between the double-planes become smaller, the line of response encounter a larger chance with an oblique angle. Reconstructed images with 1cm-spacing sever serious blurring effects. It was relieved when the plane distance become large. In a ring PET detector, to obtain images with an acceptable level of resolution uniformity, most systems restrict their FOV to 1/2-2/3 of the detector ring size. Nevertheless, in the double-plane system, the reconstructed results of the micro-Derenzo phantom demonstrated that the blurring effects could be compensated well even with a plane distance of 1 cm. Therefore, the system configure could take use of more span data and achieve a shorter scanning time without loss of resolution.

There are still some limitations exist for the algorithm. With a changeable distance between the planes, the problem is more complicated since the system response matrix is changeable for the different detector spacing. And the computation boundary is a problem should be taken into consideration. Although the openMP used to accelerate the reconstruction, the time consume is still not applicable for real-time breast imaging in biopsy. Therefore a GPU accelerate method is in expected. The positron range effect is negligible with 18F tracers. But for new radiotracer with a larger positron range, the effect should be taken into consideration.

## 5 Conclusion

A high resolution reconstruction method of a double-plane PET system was studied for breast imaging. The system is supposed with a changeable distance. A finite geometric sensitivity model was applied and resolution model derived from single gamma response was studied. And the resulting improvements in contrast recovery of small structure and noise properties in images have been demonstrated. This approach makes greater use of the high spatial resolution available with new tomographs, which is applicable for different size breast imaging. [26]